\documentclass[twocolumn,showpacs,preprintnumbers,amsmath,amssymb]{revtex4}
\usepackage{graphicx}
\usepackage{dcolumn}
\usepackage{bm}

\begin{document}

\preprint{APS/123-QED}
\title{Quantum-degenerate mixture of fermionic lithium and bosonic rubidium gases}

\author{C. Silber}
\author{S. G\"unther}
\author{C. Marzok}
\author{B. Deh}
\author{Ph.W. Courteille}
\author{C. Zimmermann}
\affiliation{Physikalisches Institut, Eberhard-Karls-Universit\"at T\"ubingen,
\\Auf der Morgenstelle 14, D-72076 T\"ubingen, Germany}

\date{\today}

\begin{abstract}
We report on the observation of sympathetic cooling of a cloud of fermionic $^6$Li atoms which are thermally coupled to evaporatively 
cooled bosonic $^{87}$Rb. Using this technique we obtain a mixture of quantum-degenerate gases, where the Rb cloud is colder than 
the critical temperature for Bose-Einstein condensation and the Li cloud colder than the Fermi temperature. From measurements of the 
thermalization velocity we 
estimate the interspecies $s$-wave triplet scattering length $|a_s|=20_{-6}^{+9}~a_B$. We found that the presence of residual rubidium 
atoms in the $|2,1\rangle$ and the $|1,-1\rangle$ Zeeman substates gives rise to important losses due to inelastic collisions. 
\end{abstract}

\pacs{05.30.Fk, 05.30.Jp, 32.80.Pj, 67.60.-g}

\maketitle

The recent realization of degenerate gases of homonuclear diatomic molecules is an important achievement for the physics of ultracold 
dilute quantum gases \cite{Herbig03,Jochim03}. The key to this spectacular success was the use of Feshbach resonances which allow for an 
adiabatic transformation from free pairs to molecules in the least bound state close to the dissociation energy \cite{Inouye98}. The 
large internal energy makes the gas unstable against collisions between the molecules, which transform the energy into kinetic energy of 
the collision partners \cite{Regal02}. The very long lifetimes of up to several $10~$s observed for dimers made of fermionic $^6$Li 
\cite{Jochim03} are explained by the fact that the atoms are only loosely bound with a binding energy of $1~\mu$K such that they still 
behave very much like individual atoms \cite{Petrov04}. Collisions are thus suppressed due to Pauli-blocking. Fermionic lithium is now 
a very promising candidate for exploring the cross-over regime between the Bardeen-Cooper-Schrieffer (BCS) model for superconductivity 
based on Cooper pairs and the regime of a Bose-Einstein condensate of composite bosons \cite{Bartenstein04}. 

In contrast to homonuclear dimers molecules made of two different atomic species exhibit a permanent electric dipole moment. The LiRb 
dimer has a very large permanent electric dipole moment of up to $4.2~$Debye \cite{Aymar05}. The intermolecular interaction in such a 
heteronuclear gas has a pronounced long-range character, which fundamentally alters its behavior at low temperatures \cite{Baranov02} 
and brings global properties of the gas into play. The geometric shape of the cloud for instance will influence the total interaction 
energy and thus the stability of the gas \cite{Santos00}. 

A possible approach to generate a polar molecular gas would use a Feshbach resonance in close analogy to the homonuclear experiments 
\cite{Herbig03,Jochim03}. Only recently first heteronuclear Feshbach resonances have been observed \cite{Stan04}. Weakly bound 
heteronuclear molecules are expected to be relatively stable against collisions if one of the two atoms is a fermion \cite{Petrov04}. 
For such molecules Pauli-blocking is at least partially effective, even though reactive collisions may reduce their lifetime 
\cite{Cvitas05}. As an alternative to Feshbach resonances, photoassociation is used to transform atomic pairs into molecules. 
Photoassociation has been extensively investigated for the homonuclear case \cite{Fioretti98}, and there are already examples of 
successful heteronuclear photoassociation \cite{Schloder01,Schloder02,Kerman04}. The advantage of photoassociation is the possibility 
to form molecules also in deeper lying bound states \cite{Sage05}, where the internuclear distance is drastically reduced and the 
molecules behave more like composite particles with a well defined quantum statistics. Furthermore, the dipole moment of heteronuclear 
molecules is largest for deeply bound states. 

Besides their importance for generating molecular gases, mixtures have remarkable physical properties which cannot be observed in pure 
quantum gases. Among them are phase separation effects \cite{Nygaard99} and collective excitations due to mutual mean field interaction 
between the mixed gases \cite{Ferrari02}. The interaction between fermions can be strongly modified in the presence of a bosonic 
background gas \cite{Heiselberg00}: Similar to phonon-induced formation of Cooper pairs in superconductors, it is expected that an 
atomic Fermi gas can be driven into a BCS transition by mediation of the Bose gas. 

The practical reason to work with mixtures is, that thermal coupling to a different species is the key to cooling a Fermi gas 
\cite{Truscott01,Schreck01} and may even be used to condense a Bose gas \cite{Modugno01}. Various mixtures of bosonic and fermionic 
alkalis are currently under investigation. In this work we report on the first studies with mixtures of fermionic $^6$Li with $^{87}$Rb. 
We demonstrate that sympathetic cooling works down to the regime of Fermi degeneracy, provided special care is taken to ensure the 
purity of the Rb cloud. Indeed, the presence of $|2,1\rangle$ or $|1,-1\rangle$ atoms leads to large inelastic Li losses at high 
densities, and their removal from the trap is a precondition to achieve Fermi degeneracy. Furthermore, we measure the interspecies 
thermalization speed and derive a value for the scattering length for heteronuclear collisions. 

\begin{figure}[ht]
		\centerline{\scalebox{0.44}{\includegraphics{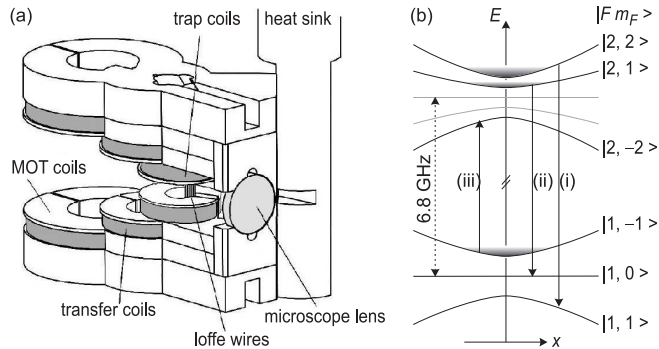}}}\caption{
			(a) Setup for simultaneous trapping of lithium and rubidium. All parts are located inside the vacuum chamber. After optical 
			cooling between the MOT-coils, both clouds are transferred into a Ioffe-Pritchard trap by adiabatic transfer with three pairs 
			of coils. In the trap the atoms are optically accessible from all six directions. Both clouds can be monitored simultaneously 
			by absorption imaging. (b) Scheme of the microwave transitions used (i) for evaporating the Rb cloud in the $|2,2\rangle$ 
			state, (ii) for removing atoms from the $|2,1\rangle$ state and (iii) for removing atoms from the $|1,-1\rangle$ state.} 
		\label{Fig1}
		\end{figure}

The scheme of our experiment is as follows: We simultaneously load $^{87}$Rb atoms from a dispenser and $^6$Li atoms from a Zeeman 
slower \cite{Schloder01} into superposed standard magneto-optical traps (MOT). From here the atoms are transferred into a magnetic 
quadrupole potential operated with the same coils as the MOT. With a spin-polarizing light pulse the Li and Rb atoms are transferred to 
the hyperfine states $|\frac{3}{2},\frac{3}{2}\rangle$ and $|2,2\rangle$, respectively. The potential is then compressed and, with an 
arrangement inspired by Ref.~\cite{Ott01}, the atoms are transferred via a second into a third quadrupole trap [cf. Fig.~\ref{Fig1}(a)]. 
Via two pairs of $0.9~$mm thick wires running parallel to the symmetry axis of the third quadrupole trap a Ioffe-Pritchard type potential 
is created \cite{Silber05}. For typical operating conditions the potential energy increases linearly with the radial distance from the 
center and quadratically with the axial distance. Only close to the center, i.e.~for temperatures of the atoms in the $\mu$K range, 
the trap is harmonic. 

The Rb cloud is cooled by forced evaporation: A microwave frequency resonantly tuned to the ground state hyperfine structure couples the 
trapped Zeeman state $|2,2\rangle$ and the untrapped $|1,1\rangle$, as shown in figure~1(b). After $15~$s of down-ramping the microwave, 
we reach the threshold to quantum degeneracy at $T_{\mathrm{c}}=620~$nK with about $N_{87}=1.2\times10^6$ atoms. For the sake of 
definiteness, here and in the following we assign the subscripts $6$ to lithium and $87$ to rubidium quantities. Typical trap frequencies 
at the end of the evaporation ramp are $\omega_x\approx\omega_y\simeq 2\pi\times206~$Hz and $\omega_z\simeq 2\pi\times50.1~$Hz, obtained 
at a bias field of $3.5~$G. Cooling down further yields almost pure condensates of $5\times10^5$ Rb atoms. The microwave is parked at 
$100~$kHz above the potential minimum to prevent heating due to glancing collision with the background gas, whose pressure is a few 
times $10^{-11}~$mbar. The condensate lifetime is about $1~$s. 
		\begin{figure}[ht]
		\centerline{\scalebox{0.55}{\includegraphics{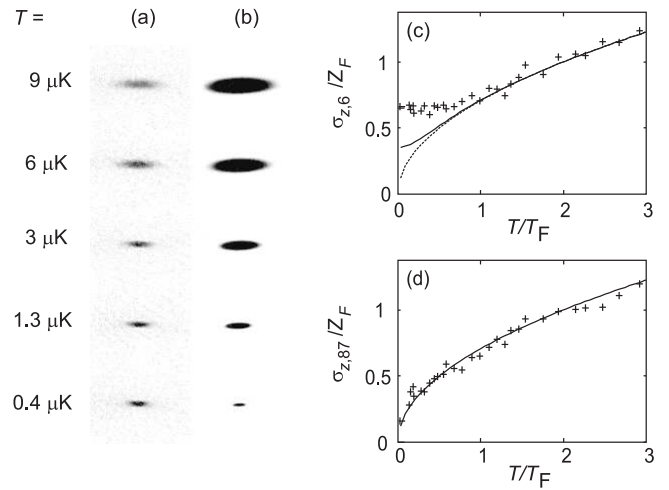}}}\caption{
			Simultaneous \textit{in-situ} absorption pictures of the Li Fermi-gas (a) and the Rb Bose-condensate (b) taken at different 
			stages of the evaporation process. Temperatures are determined from $15~$ms time-of-flight absorption images of the Rb 
			cloud. The axial (horizontal) size of the Li cloud is clearly limited to values $2\sigma_{z,6}\gtrsim 80~\mu$m. In contrast, 
			the Rb condensate shrinks at lower temperatures. From such absorption images we determine the temperature dependence of the 
			axial \textit{rms}-radii of (c) the trapped Li cloud and (d) the trapped Rb cloud. The theoretical curves show the temperature 
			dependence expected for a gas of classical particles (dotted lines) and for a gas of fermions (solid lines) with 
			$N_6=2\times10^5$ particles. The Rb and the Li data have been scaled with a common factor to rule out uncertainties in the 
			magnification of the imaging system.}
		\label{Fig2}
		\end{figure}

At the beginning of the evaporation ramp the Li cloud consists of about $2\times10^7$ atoms. In the absence of Rb atoms the lifetime of 
the Li cloud is about $170~$s. As a result of the evaporative cooling of the Rb cloud, the Li cloud is cooled sympathetically provided 
the evaporation ramp is slow enough, i.e.~in practice 25~s. We found that, if the ramp is too fast, the Li cloud thermally decouples 
from the Rb cloud. This behavior indicates a small interspecies collision rate. We also noticed dramatic losses for the Li cloud that 
will be discussed below. At the end of the evaporation ramp we let the clouds equilibrate for $1~$s before imaging them inside the trap 
[cf. Fig.~\ref{Fig2}(a,b)]. 

The Fermi temperature is reached with typically $N_6=2\times10^5~$ Li atoms \textit{above} the critical temperature for Bose-Einstein 
condensation: $T_{\mathrm{F}}=T_{\mathrm{c}}\sqrt{87/6}~[6\zeta(3)N_6/N_{87}]^{1/3}\simeq 2.4~\mu$K, where $\zeta(3)=1.202$. 
This ensures a good spatial overlap between the clouds, which is important to avoid spatial separation in the gravity field. For our 
conditions we expect a relative gravitational sag, $\Delta y=6~\mu$m, smaller than the classical \textit{rms}-radii of the clouds, 
$\sigma_{r,6}=\sigma_{r,87}\simeq 15~\mu$m. When the Rb cloud is cooled to temperatures below $T_{\mathrm{F}}$, we observe that the axial 
radius of the trapped Li cloud reaches a lower bound at values below the \textit{rms}-Fermi radius, 
$\sigma_{z,6}<Z_{\mathrm{F}}/\sqrt{2}=\sqrt{k_{\mathrm{B}}T_{\mathrm{F}}/m_{87}\omega_z^2}\simeq 62~\mu$m, but above the theoretical 
prediction, $\sigma_{z,6}>Z_{\mathrm{F}}/\sqrt{8}$ [cf. Fig.~\ref{Fig2}(c,d)]. This behavior may be explained by a joined impact of 
fermionic quantum statistics and a deceleration of sympathetic cooling as the number of Rb atoms decreases through forced evaporation. 
The weakening of the thermal coupling, which seems to have played a role in previous experiments \cite{Schreck01,Hadzibabic02}, is more 
pronounced in our case by the slow Li-Rb cross species thermalization rate. 

In the presence of Rb atoms we observe a steady decrease in the Li atom number. The decrease, which is faster at high Rb densities, 
completely dominates the time scales, and if no measure is taken to slow it down, the Li cloud disappears well before the Fermi 
temperature is reached. These Li trap losses are induced on one hand by $|1,-1\rangle$ Rb atoms, which remain in the trap because the 
initial spin polarizing pulse fails to transfer all atoms into the fully stretched state. On the other hand, we observe that 
$|2,1\rangle$ Rb atoms are produced out of the $|2,2\rangle$ cloud on a continuous basis at a rate of $G_{dip}=3.5^{\pm1.5}~$cm$^3$/s 
measured at $10~\mu$K temperature. The rate is consistent with theoretical predictions for spin relaxation due to magnetic dipole-dipole 
interactions \cite{Mies96}. The $|2,1\rangle$ atoms then inelastically collide with $|\frac{3}{2},\frac{3}{2}\rangle$ Li atoms to give 
rise to one of the following product combinations: 
$|2,2\rangle+|\frac{3}{2},\frac{1}{2}\rangle$ or $|2,2\rangle+|\frac{1}{2},\frac{1}{2}\rangle+\hbar\omega_{\mathrm{6,hfs}}$ or 
$|1,1\rangle+|\frac{3}{2},\frac{3}{2}\rangle+\hbar\omega_{\mathrm{87,hfs}}$, since the total magnetic quantum number $m_F$ is good at 
all internuclear distances \cite{Ferrari02}. Losses for dense clouds induced by populated wrong hyperfine states have been observed 
earlier \cite{Hadzibabic02,Myatt97}. 

Note that the microwave frequency used for evaporation is selective to the $|2,2\rangle$ state, so that $|2,1\rangle$ atoms remain and 
accumulate in the trap. A radiofrequency tuned between the Zeeman substates to evaporate hot $|2,1\rangle$ together with $|2,2\rangle$ 
atoms, as employed by other groups, is prohibited in our case because of its interference with the Li cloud. To eliminate the harmful 
impact of the $|2,1\rangle$ atoms, we selectively remove them from the trap. This is done by applying a magnetic field offset chosen 
high enough to energetically separate the $|2,2\rangle$ from the $|2,1\rangle$ cloud, and then tuning the microwave frequency between 
the potential minimum seen by $|2,1\rangle$ atoms and the untrapped $|1,0\rangle$ state [cf. Fig.~\ref{Fig1}(c)]. Pulse durations of 
$5~$ms have revealed long enough to empty the undesired trapped states. However, this procedure has to be repeated several times during 
the evaporation process, because the $|2,1\rangle$ state is continuously refilled. In contrast the $|1,-1\rangle$ atoms are removed from 
the trap once for all at the beginning of the evaporation ramp via irradiation of a microwave swept across the transition to the 
anti-trapped $|2,-2\rangle$ state [cf. Fig.~\ref{Fig1}(d)]. 

To determine the cross-species scattering length $a_{\mathrm{mx}}$, we have measured the thermalization time for sympathetic cooling. 
Experimentally, we make use of the fact that the thermal equilibrium between the clouds can be disturbed by evaporating the Rb cloud 
faster than the Li temperature can follow. I.e.~we rapidly cool the Rb cloud to a certain temperature and then record the evolution of 
the Li temperature as a function of time. 

The cross-species collision cross section can be extracted from the thermalization speed using the following model \cite{Delannoy01}. In 
the case of unequal collision partners about $2.7/\xi$ collisions per atom are needed for thermalization of a gas, 
$\gamma_{\mathrm{thm}}=\xi\gamma_{\mathrm{coll}}/2.7$. The reduction factor due to the mass difference of the collision partners is 
\cite{Mudrich02} $\xi=4m_6m_{87}(m_6+m_{87})^{-2}$. The collision rate 
$\gamma_{\mathrm{coll}}=\sigma_{\mathrm{mx}}\bar{v}n_{\mathrm{mx}}$ is proportional to the cross section for interspecies collisions, 
$\sigma_{\mathrm{mx}}=4\pi a_{\mathrm{mx}}^2$, the mean thermal relative velocity 
$\bar{v}=\sqrt{(8k_B/\pi)\left(T_6/m_6+T_{87}/m_{87}\right)}$, and the overlap density of the two clouds 
	\begin{equation}
	n_{\mathrm{mx}}\equiv\left(N_6^{-1}+N_{87}^{-1}\right)\int n_6(\mathbf{r})n_{87}(\mathbf{r})d^3\mathbf{r}~,\label{Eq01}
	\end{equation}
where $n_6$ and $n_{87}$ are the density distributions of the Li and the Rb clouds, respectively. The instantaneous temperature 
difference $\Delta T$ evolves according to 
	\begin{equation}
	\frac{d}{dt}(\Delta T)=-\gamma_{\mathrm{thm}}\Delta T~.\label{Eq02}
	\end{equation}
		\begin{figure}[ht]
		\centerline{\scalebox{0.48}{\includegraphics{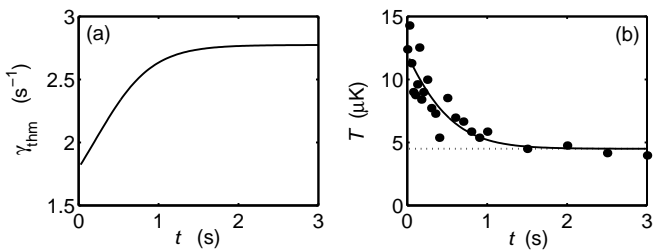}}}\caption{
			Measured and simulated evolution of the thermalization process. Both the Rb and the Li clouds are precooled to $12~\mu$K, 
			before the Rb cloud is quickly ramped down to $5~\mu$K. 
			(a) Evolution of the cross-species thermalization rate according to the model Eq.~(\ref{Eq02}). 
			(b) Measured (dots) and calculated (solid line) temperature evolution of the Li cloud. The dotted line indicates the Rb 
			temperature.}
		\label{Fig3}
		\end{figure}

Experimentally, we evaporate the Rb cloud to $12~\mu$K within $20~$s and wait for the Li cloud to follow up. We then rapidly pursue the 
evaporation ramp for the Rb cloud down to $5~\mu$K within $300~$ms. At that time we find the Li temperature still at $12~\mu$K. The 
number of Rb atoms is estimated to $(1.3\pm0.6)\times10^7$, and the Li atom number is roughly $10^5$. Starting with this situation we 
observe the gradual thermalization of the Li cloud. The microwave radiation is parked at the final frequency of the evaporation ramp, 
where it skims off those Rb atoms which are heated during the thermalization process. Because of this and because of the larger heat 
capacity of the bigger Rb cloud, its temperature remains stable, while the Li cloud reduces its temperature until thermal equilibrium 
with the Rb cloud. Figure~\ref{Fig3}(b) shows how the temperatures evolve with time. Applying the model outlined above by iterating 
equation~(\ref{Eq02}), we find that the best fit to the data is compatible with the cross species $s$-wave triplet scattering length 
$|a_{\mathrm{mx}}|=20_{-6}^{+9}~a_{\mathrm{B}}$. The small value of the interspecies scattering length explains why the sympathetic 
cooling dynamics is so slow that it decouples from the forced evaporation process of Rb. Also shown in figure~\ref{Fig3}(a) is the 
time-evolution of the calculated cross-species thermalization rate $\gamma_{\mathrm{thm}}$. Obviously, it is not constant but increases 
with time mainly because the spatial overlap improves as thermalization goes on. The measurement agrees well with a recent calculation 
of the scattering length \cite{Ouerdane04}, although the calculation should be taken with care, because it is based on inaccurately 
known interaction potentials. 

The accuracy of the measurement is limited by the uncertain number of Rb atoms. In contrast, the likewise uncertain Li atom number does 
not influence the thermalization rate, because for $N_6\ll N_{87}$ it drops out of the overlap density~(\ref{Eq01}). Furthermore, 
the simple model used to simulate the thermalization dynamics must be taken with care. In fact, it describes 
the thermalization of two coupled ensembles, each of them being separately in a thermal equilibrium characterized by a distinct 
temperature. However this assumption does not hold for the Li cloud, which can not equilibrate because of the absence of $s$-wave 
collisions. These uncertainties are accounted for by a conservative estimation of the error for the scattering length. An improved data 
analysis would require a complete numerical simulation of the thermalization dynamics \cite{Wu96}. 

In conclusion, we have observed sympathetic cooling of a cloud of fermionic lithium by an actively cooled rubidium cloud, although the 
cross-species thermalization is hindered by two facts: First of all, inelastic collisions with Rb atoms in wrong Zeeman states introduce 
important losses for the Li cloud, which quickly annihilate the cloud. And second, the very low value for the interspecies scattering 
length considerably slows down the thermalization process. By repeatedly purifying the Rb cloud and by choosing a slow cooling ramp we 
could avoid these problems and drive the Li cloud to quantum degeneracy. 

A way of manipulating the interspecies scattering length is not only desirable for controlling the thermalization dynamics, but may also 
prove essential for synthesizing heteronuclear molecules. The scattering length can be efficiently tuned near a Feshbach resonance. A 
sensible project for the near future could thus be the search for heteronuclear Feshbach resonances. Unfortunately, the interatomic 
potentials for Li-Rb collisions, and hence the location of Feshbach resonances is yet unknown. 

We acknowledge financial support from the Deutsche Forschungsgemeinschaft.

\end{document}